\documentclass[numberedappendix]{emulateapj}
\usepackage{amssymb}
\usepackage{amsmath}

\newcommand{\msol}{\,\textrm{M}_\sun}                
\newcommand{\PaperI}{Paper~I}

\newcommand{\MLV}{\Upsilon_{*\textrm{V}}}
\newcommand{\MLB}{\Upsilon_{*\textrm{B}}}
\newcommand{\MLR}{\Upsilon_{*\textrm{R}}}
\newcommand{\MLVSPS}{\Upsilon_{*\textrm{V}}^{\textrm{SPS}}}
\newcommand{\MLBSPS}{\Upsilon_{*\textrm{B}}^{\textrm{SPS}}}
\newcommand{\ML}{\Upsilon_*}

\shorttitle{The Density Profiles of Galaxy Clusters. II. Separating Luminous and Dark Matter}
\submitted{}
\begin{document}

\shortauthors{Newman et al.}
\title{The Density Profiles of Massive, Relaxed Galaxy Clusters.\\
II. Separating Luminous and Dark Matter in Cluster Cores}

\author {Andrew B. Newman\altaffilmark{1},  Tommaso Treu\altaffilmark{2},
Richard S. Ellis\altaffilmark{1}, and David J. Sand\altaffilmark{2,3}}

\altaffiltext{1}{Cahill Center for Astronomy and Astrophysics, California
Institute of Technology, MS 249-17, Pasadena, CA 91125, USA; anewman@astro.caltech.edu}
\altaffiltext{2}{Department of Physics, University of California, Santa Barbara, CA 93106, USA}
\altaffiltext{3}{Las Cumbres Observatory Global Telescope Network, Santa Barbara, CA 93117, USA}

\begin{abstract}
We present stellar and dark matter (DM) density profiles for a sample of seven massive, relaxed galaxy clusters derived from strong and weak gravitational lensing and resolved stellar kinematic observations within the centrally-located brightest cluster galaxies (BCGs). In \PaperI~of the series, we demonstrated that the \emph{total} density profile derived from these data, which span three decades in radius, is consistent with numerical \emph{DM-only} simulations at radii $\gtrsim 5-10$~kpc, despite the significant contribution of stellar material in the core. Here we decompose the inner mass profiles of these clusters into stellar and dark components. Parametrizing the DM density profile as a power law $\rho_{\textrm{DM}} \propto r^{-\beta}$ on small scales, we find a mean slope $\langle \beta \rangle = 0.50 \pm 0.10~\textrm{(random)}~{}^{+0.14}_{-0.13}~\textrm{(systematic)}$. Alternatively, cored Navarro--Frenk--White (NFW) profiles with $\langle \log r_{\textrm{core}}/\textrm{kpc}\rangle = 1.14 \pm 0.13 ^{+0.14}_{-0.22}$ provide an equally good description. These density profiles are significantly shallower than canonical NFW models at radii $\lesssim 30$~kpc, comparable to the effective radii of the BCGs. The inner DM profile is correlated with the distribution of stars in the BCG, suggesting a connection between the inner halo and the assembly of stars in the central galaxy. The stellar mass-to-light ratio inferred from lensing and stellar dynamics is consistent with that inferred using stellar population synthesis models if a Salpeter initial mass function is adopted. We compare these results to theories describing the interaction between baryons and DM in cluster cores, including adiabatic contraction models and the possible effects of galaxy mergers and active galactic nucleus feedback, and evaluate possible signatures of alternative DM candidates.
\end{abstract}


\section{Introduction}
\label{sec:intro}

The internal structure of dark matter (DM) halos is a key prediction of the cold dark matter (CDM) paradigm. Numerical simulations following the detailed structure of collisionless CDM halos \citep[e.g.,][]{NFW96,Ghigna00,Diemand05,Graham06,Gao12} generically produce a central density cusp with $\rho_{\textrm{DM}} \sim r^{-1}$, characteristic of the Navarro--Frenk--White \citep[NFW;][]{NFW96} form, probably becoming slightly shallower on very small scales \citep[e.g.,][]{Navarro10}. On the hand, simulations are only beginning to make predictions for DM halos that include baryons, which could profoundly reshape their host halos. The structure of real DM halos thus contains important information about galaxy formation, but there is currently no theoretical consensus on the magnitude or even sign of these baryonic effects, particularly over a wide range in mass. Additionally, the microphysics of the unknown DM particle could become important in the densest regions, and the inner structure of DM halos may therefore provide valuable indirect clues to its nature \citep[e.g.,][]{Spergel00,Abazajian01,Kaplinghat05,Peter10}.

Given the current uncertainty, observations are clearly in a good position to guide theoretical efforts. However, measurements of DM mass profiles are extremely challenging and are usually limited by confusion with baryons, the small dynamic range of the observations, and degeneracies that are inherent to individual mass probes (e.g., velocity anisotropy). Clusters of galaxies are promising locations to make progress. Accurate mass measures are available via many independent observational probes, especially gravitational lensing and X-ray emission (see \citealt{Allen11,Kneib11} for reviews, and references in \PaperI). As we have shown \citep{N09,N11}, combining stellar kinematics with strong and weak gravitational lensing yields constraints over three decades in radius. This is comparable to the best simulations and is thus suitable for detailed comparison of the DM profile shape if the baryonic mass can be constrained. On small scales in relaxed clusters, the latter is dominated by stars in the central brightest cluster galaxy (BCG).

\citet{Sand02,Sand04} demonstrated the utility of combining resolved stellar kinematics with strong lensing to constrain two-component mass models, i.e., the BCG stars and DM halo separately (see \citealt{MiraldaEscude95,Natarajan96}). \citet{Sand04} studied six clusters and inferred a mean $\langle \beta \rangle = 0.52 \pm 0.05$, where $\rho_{\textrm{DM}} \propto r^{-\beta}$, significantly shallower than an NFW cusp having $\beta = 1$. They further noted possible variation in $\beta$ from cluster to cluster. \citet{Sand08} improved on this analysis for two clusters (MS2137 and A383) by relaxing the assumption of axial symmetry in the lensing analysis, instead conducting a full two-dimensional study. They found this did not alter their earlier findings, but noted that the inferred DM slope $\beta$ is sensitive to the adopted scale radius $r_s$, which could only be constrained by additional mass probes at larger radii. This was implemented by \citet{N09} in A611 through the addition of weak lensing data. In \citet{N11}, we further extended the methodology in A383 by constraining the role of projection effects (i.e., line-of-sight (l.o.s.) ellipticity; \citealt{Gavazzi05}) via a comparison of X-ray and lensing data. We also presented a radially-extended velocity dispersion profile measured in a very deep spectroscopic observation. In both A611 and A383, we confirmed a shallow inner DM cusp with $\beta < 0.3$ (68\% CL) and $\beta = 0.59^{+0.30}_{-0.35}$, respectively.\footnote{In the present paper we present a significantly revised measurement for A611; see Section~\ref{sec:dmresults}.}

In \PaperI~of the present series, we presented strong and weak lensing and stellar kinematic data for a sample of seven massive ($M_{200} =  0.4 - 2 \times 10^{15} \msol$), relaxed galaxy clusters at $z = 0.19 - 0.31$. This built upon our earlier papers by enlarging the sample of clusters with the highest-quality data: weak lensing measured using deep multi-color imaging, primarily from the Subaru telescope, extended stellar kinematic profiles in the BCG obtained primarily at the Keck telescopes, and multiply-imaged sources located in \emph{Hubble Space Telescope} (\emph{HST}) imaging (25 strongly lensed sources in total, of which 21 have spectroscopic redshifts). We showed that the \emph{total} inner density profile is remarkably well-described by numerical simulations containing \emph{only CDM} at radii $\gtrsim 5-10$~kpc, despite the significant contribution of stellar mass on these scales.

Here we extend \PaperI~by dissecting the stellar and DM contributions, using improved versions of the techniques developed in our earlier papers. We first show how the mass content of the BCGs can be constrained using information from the entire sample. We then isolate the DM density profiles and quantify their behavior on small scales. We show that the DM profiles become shallower than NFW models within $\approx 30$~kpc, roughly the typical effective radius of the BCGs. Furthermore, the inner DM density profiles exhibit likely variation from cluster to cluster, and this variation is correlated with the properties of the BCG. Finally, we interpret our results in the context of the recent theoretical literature, focusing on the interactions between baryons and DM in galaxy clusters and the possibility that cores in galaxy clusters are imprints of DM particle physics.

\begin{figure*}
\centering
\includegraphics[width=0.9\linewidth]{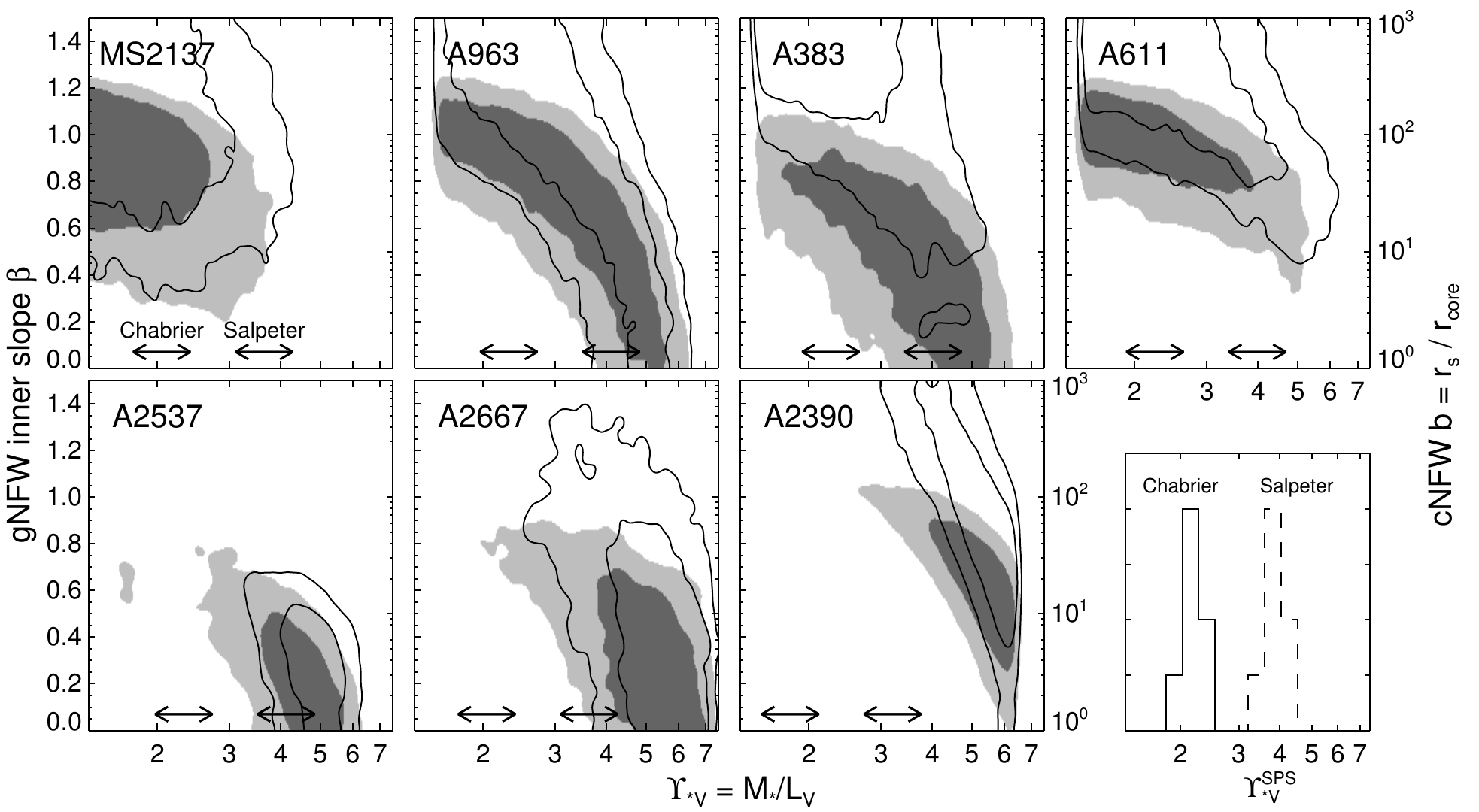}
\caption{Degeneracy between $\MLV$ and the DM inner density profile when the halo is parameterized with a gNFW (left axis, filled contours) or cNFW (right axis, lines) model. The 68\% and 95\% confidence regions are shown. Arrows at the bottom of each panel show estimates $\MLVSPS$ derived from SPS fits to the broadband colors of the BCGs, spanning the 68\% confidence interval, when adopting Chabrier and Salpeter IMFs; the distribution of these is shown in the bottom right panel. The permitted range in $\MLV$ is set proportionally to $\MLVSPS$ and therefore varies slightly from cluster to cluster (see Section~7 of Paper I).\label{fig:degen}}
\end{figure*}

Throughout we adopt a $\Lambda$CDM cosmology with $\Omega_m = 0.3$, $\Omega_{\Lambda} = 0.7$, and $H_0 = 70$~km~s${}^{-1}$~Mpc${}^{-1}$. Error bars and upper limits encompass the 68\% confidence interval. When pairs of errors are quoted, they refer to the random and systematic components, respectively.

\section{Data and Modeling}
\label{sec:datamodel}

Whereas the \emph{total} density profiles were studied in \PaperI, the goal of this paper is to use our two-component fits to separate the stellar and dark mass contributions in the cluster cores. All aspects of the data and modeling were discussed extensively in \PaperI~(Section 7). Here we provide a summary of the features most relevant for this paper. Firstly, the stellar mass in the BCG is modeled based on fits to the surface luminosity in \emph{HST} imaging. A uniform stellar mass-to-light ratio $\ML$ is assumed within each BCG. As discussed in \PaperI~(Sections 5.1 and 9.3), this is justified by the mild or null color gradients observed over the relevant radial interval. Non-BCG cluster galaxies -- relevant as perturbations in the strong lens model -- are included via scaling relations based on the fundamental plane (\PaperI, Section 7).

Secondly, the cluster-scale smooth DM halo is parameterized using either a generalized NFW (gNFW) model
\begin{equation}
\rho_{\textrm{DM}}(r) = \frac{\rho_s}{(r/r_s)^{\beta} (1 + r/r_s)^{3-\beta}} \label{eqn:gnfw}
\end{equation}
or a cored NFW (cNFW) model with
\begin{equation}
\rho_{\textrm{DM}}(r) = \frac{b \rho_s}{(1 + b r/r_s)(1+r/r_s)^2}. \label{eqn:cnfw}
\end{equation}
Both models have the same large scale behavior ($\rho_{\textrm{DM}}\propto r^{-3}$), but the gNFW model contains a central power-law cusp with $d \log \rho_{\textrm{DM}} / d \log r \rightarrow -\beta$ as $r \rightarrow 0$, while the cNFW model asymptotes to a constant-density core within a characteristic radius $r_{\textrm{core}}=r_s/b$. Both models contain the NFW profile in the limits $\beta = 1$ and $r_{\textrm{core}} \rightarrow 0$ and therefore allow us to explore deviations from canonical CDM halos in the central regions. Broad, uninformative priors are placed on the halo parameters ($\rho_s, r_s,$ and $\beta$ or $b$) and $\ML$ (\PaperI, Table 7).

Based on the close alignment between the optical centers of the BCGs and both the X-ray centroids (typically separated by $\simeq 3$~kpc, comparable to the measurement errors) and the lensing-derived centers of mass, we fix the center of the halo to that of the BCG (\PaperI, Sections 2 and 7.3). Furthermore, mass estimates derived from lensing agree well with independent X-ray observations, which constrains the l.o.s. ellipticity of the halo to be mild in all clusters except A383 \citep[see Paper~I, Section 8 and][]{N11}. 

We do not specifically distinguish the hot gas in the intracluster medium (ICM), which is thus implicitly included in the halo in our models. Since the distribution of the ICM is similar to that of the halo and comprises only a $\simeq 10\%$ mass fraction \citep[e.g.,][]{Allen04}, subtracting the ICM to isolate the DM has very little effect on the \emph{slope} of the density profile ($\Delta d \log \rho / d \log r \lesssim 0.05$; \citealt{N09,SommerLarsen10}), which is the main focus of this paper.

These models are constrained by three data sets. Firstly, the mass on scales of $\simeq 100$~kpc to 3~Mpc is constrained using gravitational shear (weak lensing) measured in deep, multi-color images primarily from the Subaru telescope. Secondly, the angular positions and redshifts of background galaxies that are strongly lensed by the clusters precisely constrain the mass from $\simeq 20$ to $100$~kpc, varying from cluster to cluster. In total we located 25 multiply-imaged sources, of which 21 have spectroscopic redshifts (7 were first presented in \PaperI). Finally, the most unique aspect of our analysis is the inclusion of spatially resolved stellar kinematics within the BCGs. These measures are derived from long-slit spectra primarily obtained at the Keck telescopes. The stellar kinematic data typically extend to $R \approx 10-20$~kpc and display a very homogenous shape that rises with radius, indicating a rising total mass-to-light ratio as expected at the centers of massive clusters. As demonstrated in \PaperI~(Section 9), the mass models provide good fits to the full range of data.

For the purpose of distinguishing dark and stellar mass, the most important physical assumptions are that stellar mass follows light, as justified above, and that the DM halo is adequately described by a gNFW- or cNFW-like profile. The precise parametric form is not as critical as the presumption that the DM density turns over smoothly at small radii -- either to a power-law cusp in the gNFW case, or to a constant density in the cNFW models -- without a sharp upturn on small scales. This is reasonable: by design these profiles describe pure CDM halos in the appropriate limits, and although the effects of adding baryons are uncertain, adiabatic contraction prescriptions \citep{Gnedin04,Gnedin11} predict DM profiles that are well-fit by gNFW models over the relevant range of radii when applied to halos and BCGs representative of our sample. Due to the density and radial extent of observational constraints (extended stellar kinematic profiles, strongly lensed galaxies usually at multiple redshifts, weak lensing), we emphasize that we are able to consider quite general families of DM halos in each cluster.


\section{Separating Luminous and Dark Mass: The Role of the Stellar Mass-to-Light Ratio}

\begin{figure}
\centering
\includegraphics[width=0.75\linewidth]{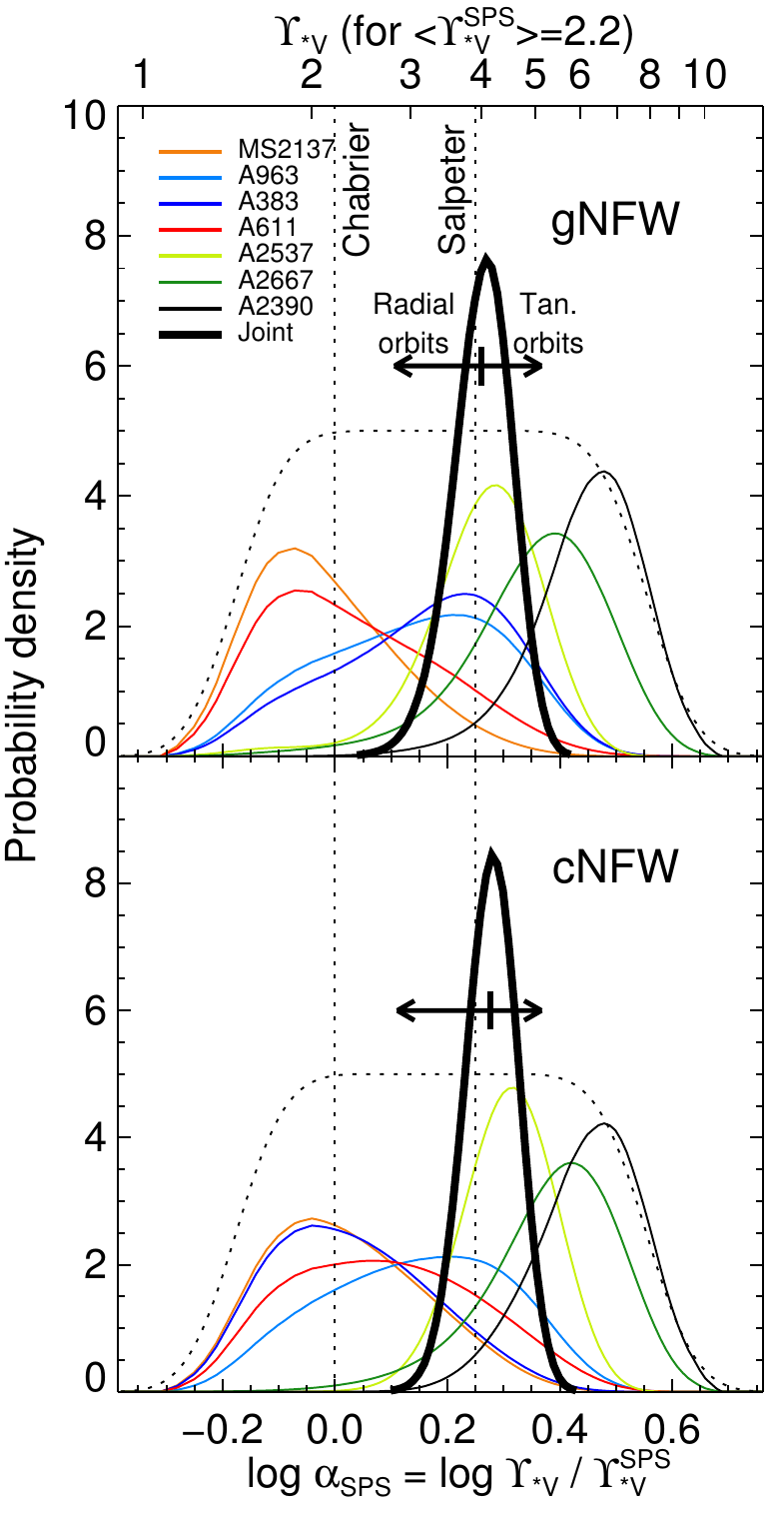}
\caption{Probability densities for $\log \alpha_{\textrm{SPS}}$, which parameterizes the stellar mass-to-light ratio $\MLV$ relative to SPS models (Equation~\ref{eqn:alphaSPS}), are shown for each cluster (thin lines) and jointly for the entire sample (thick). The dotted curve shows the effective prior, composed of the flat prior on $\log \MLV$ convolved with a Gaussian uncertainty on $\MLVSPS$ of $\sigma_{\textrm{SPS}}=0.07$~dex. Arrows indicate the effect of adopting mildly anisotropic orbits with $\beta_{\textrm{aniso}} = \pm 0.2$.\label{fig:alphaIMF}}
\end{figure}

In individual clusters there is a degeneracy between the stellar mass-to-light ratio $\MLV = M_*/L_{\textrm{V}}$ and the inner DM slope. This is illustrated in Figure~\ref{fig:degen}, which shows results for the mass models summarized in Section~\ref{sec:datamodel} and derived in \PaperI~(Section~9). This degeneracy is expected, since stellar mass in the BCG can be traded against DM. Owing to the multiplicity of constraints described above, particularly kinematic measurements at small radii in the stellar-dominated regime, the model degeneracy is not complete, and each cluster does carry information on both $\ML$ and $\beta$ or $b$.

It is already evident in Figure~\ref{fig:degen} that most of the clusters in our sample prefer a DM inner slope that is shallower than an NFW profile (i.e., $\beta < 1$), consistent with our previous findings \citep{Sand02,Sand04,Sand08,N09,N11}. However, it is also clear that the precision of the constraints on the inner slope could be increased if additional information regarding $\ML$ is available. Indeed, most clusters are consistent with a wide range of $\ML$ when viewed in isolation, due to the uncertainty arising from the degeneracy described above. Furthermore, the figure suggests a possible variation from cluster to cluster in the DM inner slope, but this conclusion may be contingent upon substantial variations in $\ML$ as well. We do not have strong a priori expectations about the possible variation from cluster to cluster in the DM inner slope, particularly recalling the uncertain role of baryons in theoretical predictions. There are, however, several strong reasons to believe that the true physical variation in $\ML$ within our sample is small.

Firstly, Figure~\ref{fig:degen} shows estimates of the stellar mass-to-light ratio $\MLVSPS$ derived by fitting stellar population synthesis (SPS) models to the broadband colors of the BCGs (see \PaperI, Section 5.2). Currently, SPS models cannot predict absolute masses more accurately than a factor of $\simeq 2$, primarily due to the unknown stellar initial mass function (IMF), which we discuss further in Section~\ref{sec:MLcompare}. On the other hand, \emph{relative} stellar masses are more robust, especially within a homogeneous galaxy population. As the bottom right panel of Figure~\ref{fig:degen} demonstrates, the range in $\MLVSPS$ within our sample at a fixed IMF is small. Assuming a \citet{Chabrier03} IMF, the median $\langle \MLVSPS \rangle = 2.2$; the full range is only $1.80-2.32$, and the rms scatter is $9\%$.\footnote{Throughout, $L_{\textrm{V}}$ and $\MLV$ refer to the observed luminosity, including any internal reddening from dust within the BCG. If we removed the reddening to obtain the intrinsic $L_{\textrm{V}}$ and $\MLV$ of the stellar populations, their scatter would increase. (Reddening is indicated only in cool core clusters hosting some current star formation.) However, the SPS \emph{stellar mass} estimates, which are significant for our analysis, are much more robust.}

Secondly, the rms dispersion in the absolute luminosities $L_{\textrm{V}}$ of the BCGs in our sample is only 0.1~dex. This small variation is consistent with previous studies of BCGs as ``standard candles'' with uniform luminosities and colors \citep[e.g.,][]{Sandage72,Postman95,Collins98,Bernardi07}. Finally, the environments of the BCGs are the same: by construction they are all central galaxies in massive clusters, and their central velocity dispersions are comparable. It would be very surprising if this uniformity in luminosity and $\MLVSPS$, which are thought to derive from a similar assembly history, were the result of a conspiracy that masks larger variations in stellar mass. Instead, based on these physical similarities, it is very likely that the BCGs in our sample have similar stellar masses and $\MLV$. As we discuss in Section~\ref{sec:MLcompare}, this is further supported by recent, independent studies.

With the well-motivated assumption that the BCGs in our sample have a similar $\MLV$, we can use the full sample of seven clusters to \emph{jointly} constrain its value, thereby improving the precision and robustness of our measurements of the DM profile. Before embarking on this,  we consider how to handle the small variations in $\MLV$ that we do anticipate, despite the overall similarity. The sample spans a redshift range $z = 0.19 - 0.31$, so some mild passive evolution is expected. Additionally, the BCGs with the lowest $\MLVSPS$ estimates show optical emission lines and far-infrared photometry indicative of ongoing star formation (although it involves a small fraction of the stellar mass; see \PaperI). These BCGs reside in the cool core clusters, consistent with earlier studies \citep{Bildfell08,Loubser09,Sanderson09a}. 

Therefore, a more precise technique is to define $\ML$ for each cluster \emph{relative to the SPS measurement}:
\begin{equation}
\log \alpha_{\textrm{SPS}} = \log \MLV / \MLVSPS.
\label{eqn:alphaSPS}
\end{equation}
We can then use the full cluster sample to constrain $\langle \log \alpha_{\textrm{SPS}} \rangle$, which parameterizes a common, systematic offset from photometrically-derived stellar mass-to-light ratios. As describe in Section~\ref{sec:MLcompare}, the most probable source for large systematic offsets is an IMF that differs from that assumed in the SPS models: in this case, that of Chabrier. However, our analysis does \emph{not} depend on the physical origin of the offset, only that is it common among our BCGs. 
Since the variation in $\MLVSPS$ is small compared to the range of $\MLV$ explored in our fits (25\% versus a factor of 5.3), this approach is not radically different from assuming a common $\MLV$. However, it improves on that assumption by making use of SPS models to adjust for small differences in $\MLV$ arising from age and dust, while making no assumption on the validity of their absolute mass scale.

\subsection{Constraining the stellar mass scale}
\label{sec:massscale}

Figure~\ref{fig:alphaIMF} shows the probability distribution for $\log \alpha_{\textrm{SPS}}$ derived in each cluster. The uncertainty in $\log \alpha_{\textrm{SPS}}$ arises from two sources: that in the $\MLV$ derived from dynamics and lensing, and the uncertainty in $\MLVSPS$ arising from random photometric errors. In \PaperI~we estimated the latter as $\sigma_{\textrm{SPS}} = 0.07$~dex. Thus, the probability distributions for $\log \alpha_{\textrm{SPS}}$ are derived by broadening those for $\log \MLV$ by a Gaussian with a dispersion of $\sigma_{\textrm{SPS}}$. 

We have argued that there are strong a priori reasons to expect that $\alpha_{\textrm{SPS}}$ is uniform across our sample of BCGs. Using the probability distributions in Figure~\ref{fig:alphaIMF}, we can ask whether the lensing and kinematic data are indeed consistent with this assumption. One way to quantify this is to suppose that the true distribution of $\log \alpha_{\textrm{SPS}}$ is Gaussian and infer its intrinsic dispersion $\sigma_{\log \alpha}$. The formalism for inferring the probability distribution $P(\sigma_{\log \alpha})$ was discussed in \PaperI~(Section~9, and see \citealt{Bolton12}). The preference for non-zero intrinsic scatter can then be assessed by
\begin{equation}
\Delta P = \sqrt{2 \ln[P(\sigma_{\log \alpha} = \sigma_{\textrm{peak}}) / P(\sigma_{\log \alpha} = 0)]},
\label{eqn:DeltaP}
\end{equation}
where $\sigma_{\textrm{peak}}$ is the location of the maximum of $P(\sigma_{\log \alpha})$. For a Gaussian distribution, $\Delta P$ is the number of standard deviations from the mean. We find $\Delta P = 0.85$, i.e., a $<1\sigma$ preference for intrinsic scatter. Thus, the lensing and kinematic data are consistent with (although they alone cannot prove) our assumption that there is little intrinsic variation in $\alpha_{\textrm{SPS}}$ within our sample.

With the physically-motivated assumption that $\alpha_{\textrm{SPS}}$ is the same for each BCG, we can constrain its common value simply by multiplying the seven independent probability distributions. The results are shown by the thick curves in Figure~\ref{fig:alphaIMF}. Very similar values of $\log \alpha_{\textrm{SPS}}=0.28\pm 0.05$ and $0.26\pm 0.05$ are derived using the gNFW and cNFW models, respectively, demonstrating that these results do not strongly depend on the exact halo model. Given the closeness of these results, in the following analysis we adopt $\log \alpha_{\textrm{SPS}} = 0.27 \pm 0.05$. Despite marginalizing over fairly general parameterizations of the DM profile, we are able to obtain informative results due to the high density of observational constraints and the sample size. 

Taking the median $\langle \MLVSPS \rangle = 2.2$, we find that $\log \alpha_{\textrm{SPS}} = 0.27$ corresponds to $\MLV = 4.1$. In Section~\ref{sec:betasys}, we describe sources of systematic uncertainty leading to a final estimate $\log \alpha_{\textrm{SPS}} = 0.27 \pm 0.05 {}^{+0.10}_{-0.16}$. In Section~\ref{sec:MLcompare}, we discuss the physical implications of this result and compare to the recent literature on the stellar mass-to-light ratio and IMF in early-type galaxies.

\begin{figure*}
\centering
\includegraphics[width=\linewidth]{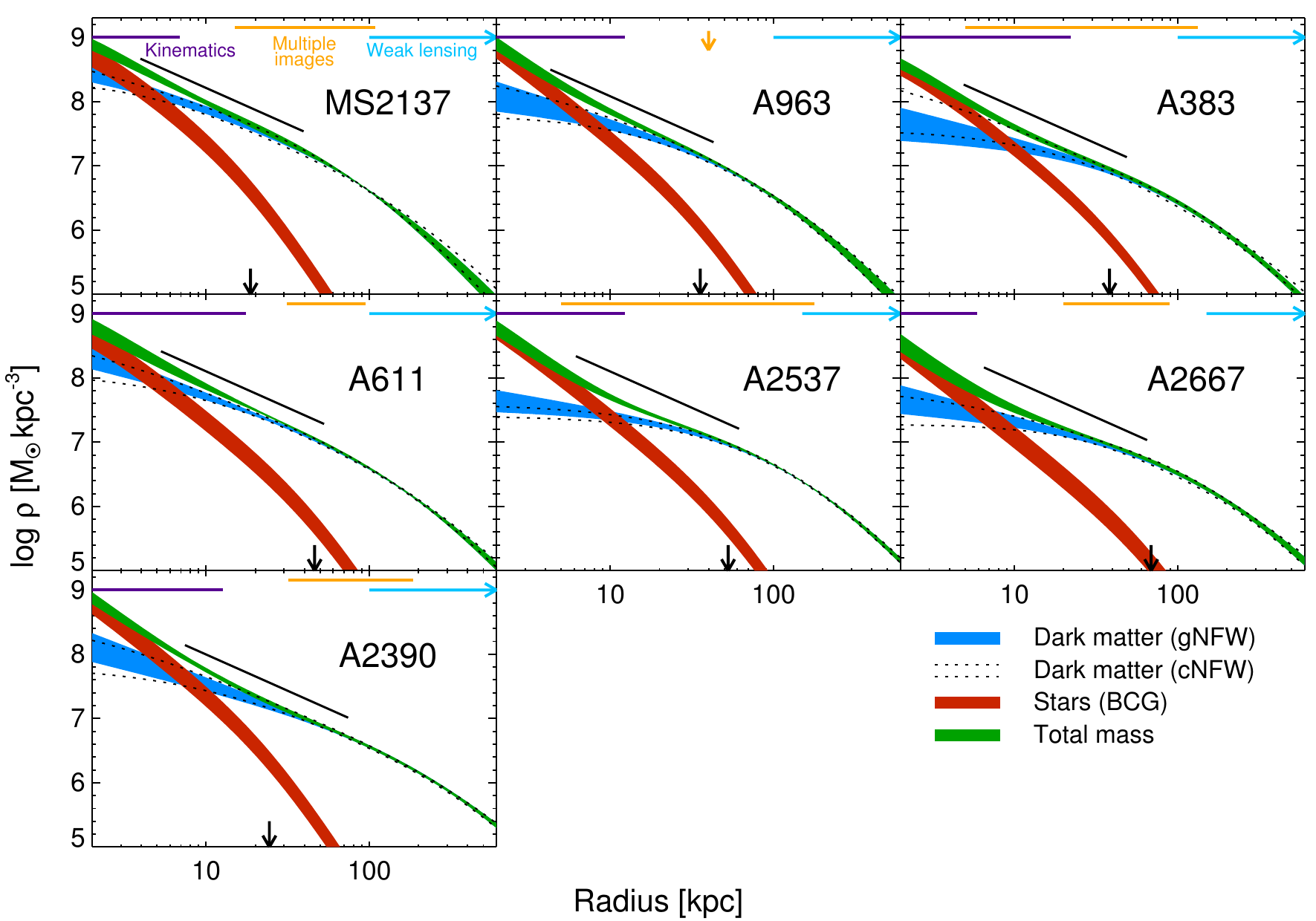}
\caption{Radial density profiles for the DM halo, stars in the BCG, and their total in each cluster, adopting the joint constraint on $\log \alpha_{\textrm{IMF}}$ from Figure~\ref{fig:alphaIMF}. The spatial extent of each data set is indicated at the top of each panel. The black line segment spans $r/r_{200}=0.003-0.03$ and has the slope $\rho \propto r^{-1.13}$ that is the average of clusters in the Phoenix DM-only simulations \citep{Gao12} over the same interval (\PaperI, Section 10). The width of the bands indicates the uncertainty, including both $1\sigma$ random uncertainties for isotropic models and a systematic component estimated by taking $\beta_{\textrm{aniso}}=\pm0.2$ (Section~\ref{sec:betasys}). Arrows at the bottom of each panel indicate three-dimensional half-light radius $r_h$ of the BCG. \label{fig:dens}}
\end{figure*}

\section{The Inner Dark Matter Density Profile}

We now turn to the the inner DM density profiles. In our earlier papers, we studied the inner DM density slope $\beta$ by marginalizing over the uncertainty in $\ML$ separately in each cluster. With the benefit of a larger sample with improved data, we have now combined constraints from seven clusters to arrive at a joint measurement of the stellar mass scale $\alpha_{\textrm{SPS}}$ (Section~\ref{sec:massscale}). Incorporating this information, we can now conduct our analysis in a more physically consistent way that recognizes the homogeneity of the BCGs, as well as further reducing the remaining degeneracies between dark and stellar mass.

Technically, we implement the joint constraint on $\log \alpha_{\textrm{SPS}}$ via importance sampling \citep[e.g.,][]{Lewis02}, reweighting the Markov chain samples to effectively convert our flat prior on $\log \alpha_{\textrm{SPS}}$ to a Gaussian with mean $\langle \log \alpha_{\textrm{SPS}} \rangle = 0.27$ and dispersion $\sigma = (\sigma_{\alpha}^2 + \sigma_{\textrm{SPS}}^2)^{1/2} = 0.09$. Here $\sigma_{\alpha} = 0.05$~dex is the uncertainty in $\langle \log \alpha_{\textrm{SPS}} \rangle$, and $\sigma_{\textrm{SPS}} = 0.07$~dex is the random error in $\MLVSPS$ for each BCG. The latter accounts for the fact that $\alpha_{\textrm{SPS}}$ refers to a systematic offset from SPS-based mass estimates, but random errors due to photometric noise remain in each cluster.\footnote{This estimate of $\sigma_{\textrm{SPS}}$ may be conservative, given that the dispersion in $\MLVSPS$ measurements among the BCGs is smaller, and $\chi^2/\textrm{dof} \leq 1$ in the SPS model fits. Thus, in practice we are likely allowing for some mild intrinsic variation in $\alpha_{\textrm{SPS}}$.}

\subsection{Dark and stellar mass profiles}

Figure~\ref{fig:dens} shows the resulting spherically-averaged density profiles for the DM halo, BCG stars, and their sum.
The results based on gNFW and cNFW models are again quite similar, showing that the choice of parameterization does not strongly affect the derived density profiles. We do not detect an overall preference for one model over the other: the ratio of the total Bayesian evidence is consistent with unity.\footnote{In Paper I we found that the evidence ratio mildly favored the cNFW models when taking a uniform prior on $\log \alpha_{\textrm{SPS}}$. When the joint constraint derived in this paper is taken as a prior, the evidence ratio is consistent with unity ($\ln E_{\textrm{gNFW}} / E_{\textrm{cNFW}} = -0.8 \pm 3.2$).}

The black line segment in each panel spans $r/r_{200} = 0.003 - 0.03$, which is the interval over which the total density slope $\gamma_{\textrm{tot}}$ was defined in \PaperI. Its slope $r^{-1.13}$ is the average measured in CDM-only cluster simulations from the Phoenix project \citep{Gao12}. As quantified in \PaperI, the stars and DM sum to produce a slope very close to CDM-only simulations over this interval.

Now we can see that both stars and DM contribute significantly to the mass in this regime: stars dominate the density in the inner radius, while virtually all the mass is DM at the outer radius. This demonstrates a tight coordination between the inner DM profile and the distribution of stars: the NFW-like density slope is not a property of the DM halo or the BCG alone, but of their sum.
As noted in \PaperI, at yet smaller radii $r \lesssim 5-10$~kpc where stars are dominant -- well within the mean effective radius $\langle R_e \rangle = 30$~kpc -- the total density profile generally steepens.

As expected if the total density is NFW-like, the DM profiles become shallower only on scales where the BCG contributes significantly, roughly within $R_e$. As we describe in Section~\ref{sec:compare}, our results thus do not conflict with other studies that claim the DM alone follows an NFW profile but are confined to $r \gtrsim R_e$. The stellar mass density in our models reaches that of the DM at a median radius of $\langle r \rangle = 7$~kpc. In terms of enclosed mass, equality occurs at $\langle r \rangle = 12$~kpc. Within 5~kpc the median DM fraction is $\langle f_{\textrm{DM}}\rangle=25\%$, similar to massive field ellipticals \citep[e.g.,][]{Auger10b}, but within their three-dimensional half-light radii $r_h$ the BCGs are far more DM-dominated: $\langle f_{\textrm{DM}} \rangle=80\%$. 

\begin{deluxetable}{lcc}
\tablewidth{\linewidth}
\tablecolumns{3}
\tablecaption{Parameters Describing the Inner DM Profile}
\tablehead{\colhead{Cluster} & \colhead{$\beta$ (gNFW)} & \colhead{$\log r_{\textrm{core}}$/kpc (cNFW)}}
\startdata
MS2137 & $0.65^{+0.23}_{-0.30}$ & $0.45^{+0.38}_{-0.48}$ \\
A963 & $0.50^{+0.27}_{-0.30}$ & $0.87^{+0.61}_{-0.71}$ \\
A383 & $0.37^{+0.25}_{-0.23}$ & $0.37^{+0.72}_{-0.64}$ \\
A611 & $0.79^{+0.14}_{-0.19}$ & $0.47^{+0.39}_{-0.50}$ \\
A2537 & $0.23^{+0.18}_{-0.16}$ & $1.67^{+0.24}_{-0.23}$ \\
A2667 & $0.42^{+0.23}_{-0.25}$ & $1.29^{+0.49}_{-0.49}$ \\
A2390 & $0.82^{+0.13}_{-0.18}$ & $0.30^{+0.53}_{-0.34}$ \\
\cutinhead{Ensemble average}
\textbf{All clusters} & $0.50\pm0.13$ & $1.14\pm0.13$ \\
$\beta_{\textrm{aniso}} = +0.2$ & $0.38^{+0.09}_{-0.07}$ & $1.11^{+0.14}_{-0.10} $ \\
$\beta_{\textrm{aniso}} = -0.2$ & $0.64^{+0.05}_{-0.09}$ & $0.96^{+0.24}_{-0.11} $ \\
Separate $\alpha_{\textrm{SPS}}$ & $0.62\pm0.14$ & $1.09^{+0.12}_{-0.21}$
\enddata
\tablecomments{Median parameters are shown, obtained after weighting
samples to incorporate our joint constraint on $\alpha_{\textrm{SPS}}$, as described in the text. Error bars encompass the $16-84$th percentiles and account for random uncertainties only; see Section~\ref{sec:betasys} for a discussion of systematic errors. Results are shown for individual clusters (top) and for the ensemble mean (bottom), including for several alternative assumptions described in Section~\ref{sec:betasys}.\label{tab:beta}}
\end{deluxetable}

\subsection{Inner DM density slopes and core radii}
\label{sec:innerslopes}

\begin{figure*}
\centering
\includegraphics[width=0.45\linewidth]{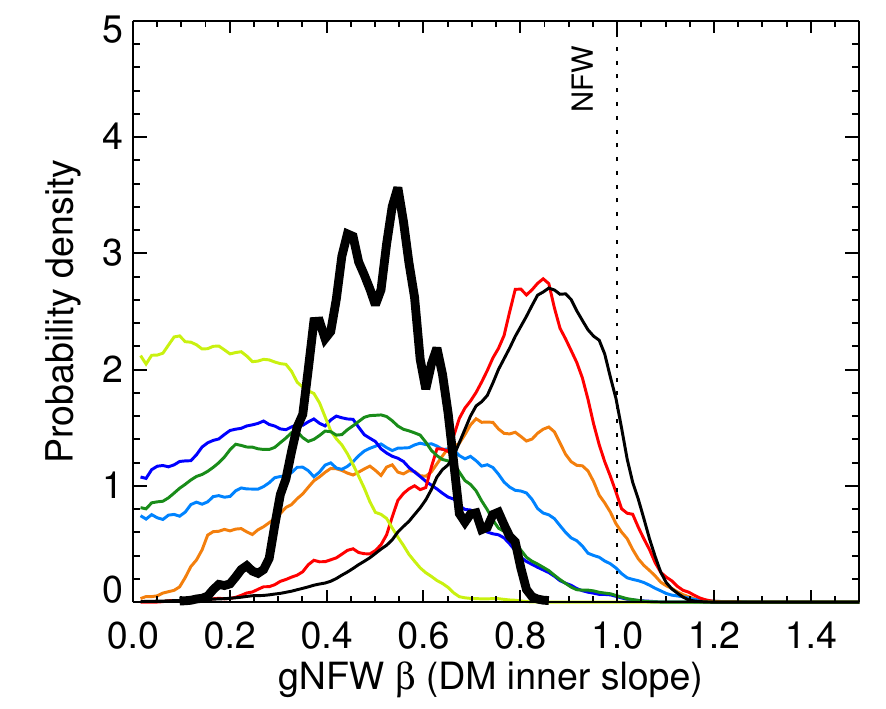} \hspace{0.5cm}
\includegraphics[width=0.45\linewidth]{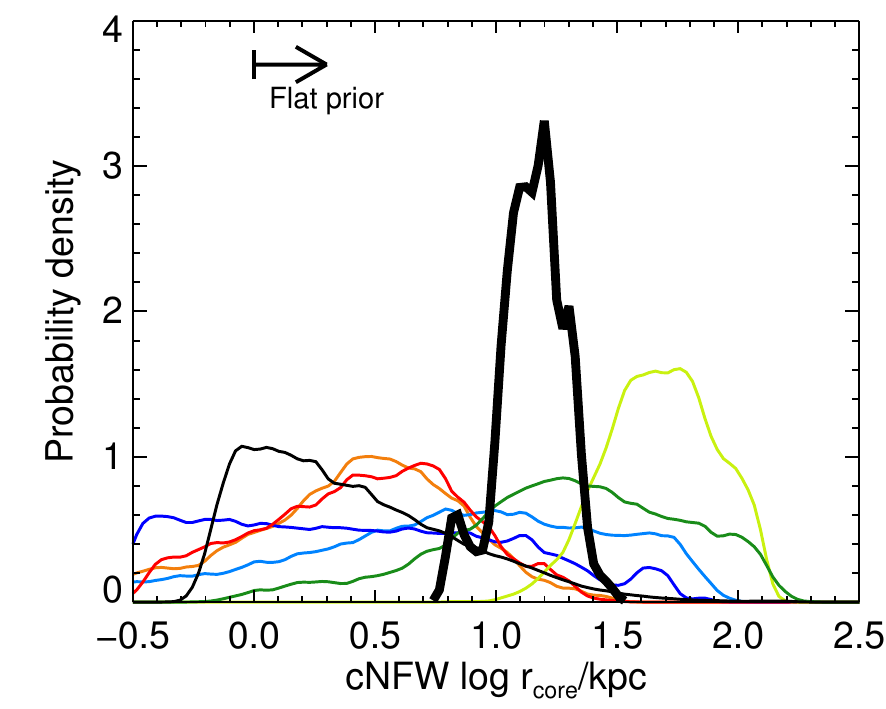}
\caption{Marginalized posterior probability densities for the gNFW inner slope $\beta$ (left) and the cNFW core radius (right). Thick black lines show joint constraints on the mean. Colors follow Figure~\ref{fig:alphaIMF}. Each curve is normalized to integrate to unity. Our adopted priors (\PaperI, Table~7) are flat in $\log r_s$ and $b$ and therefore not in $\log r_{\textrm{core}} = \log r_s / b$ over the whole plotted range. The arrow in the right panel indicates the region over which the effective prior on $\log r_{\textrm{core}}$ is flat. \label{fig:DMparams}}
\end{figure*}

Figure~\ref{fig:DMparams} shows the probability distributions for $\beta$ (gNFW) and $r_{\textrm{core}}$ (cNFW) obtained by marginalizing over the other parameters, again weighting the samples to incorporate our joint constraint on $\alpha_{\textrm{SPS}}$. Results for the individual clusters are listed in Table~\ref{tab:beta}. Every cluster prefers $\beta < 1$, i.e., an inner slope shallower than an NFW model. Thick black lines shows constraints on the mean: $\langle \beta \rangle = 0.50 \pm 0.13$ and $\langle \log r_{\textrm{core}}/\textrm{kpc} \rangle = 1.14 \pm 0.13$; the method for deriving these is outlined in the Appendix. We note that while the typical $r_{\textrm{core}} \approx 14$~kpc is small, the cNFW profile turns over rather slowly at small radii. Thus, while $r_{\textrm{core}}$ is the radius where the density falls to half of the corresponding NFW profile, significant deviations extend to $r \simeq (3-4) r_{\textrm{core}}$.

\begin{figure}
\centering
\includegraphics[width=0.9\linewidth]{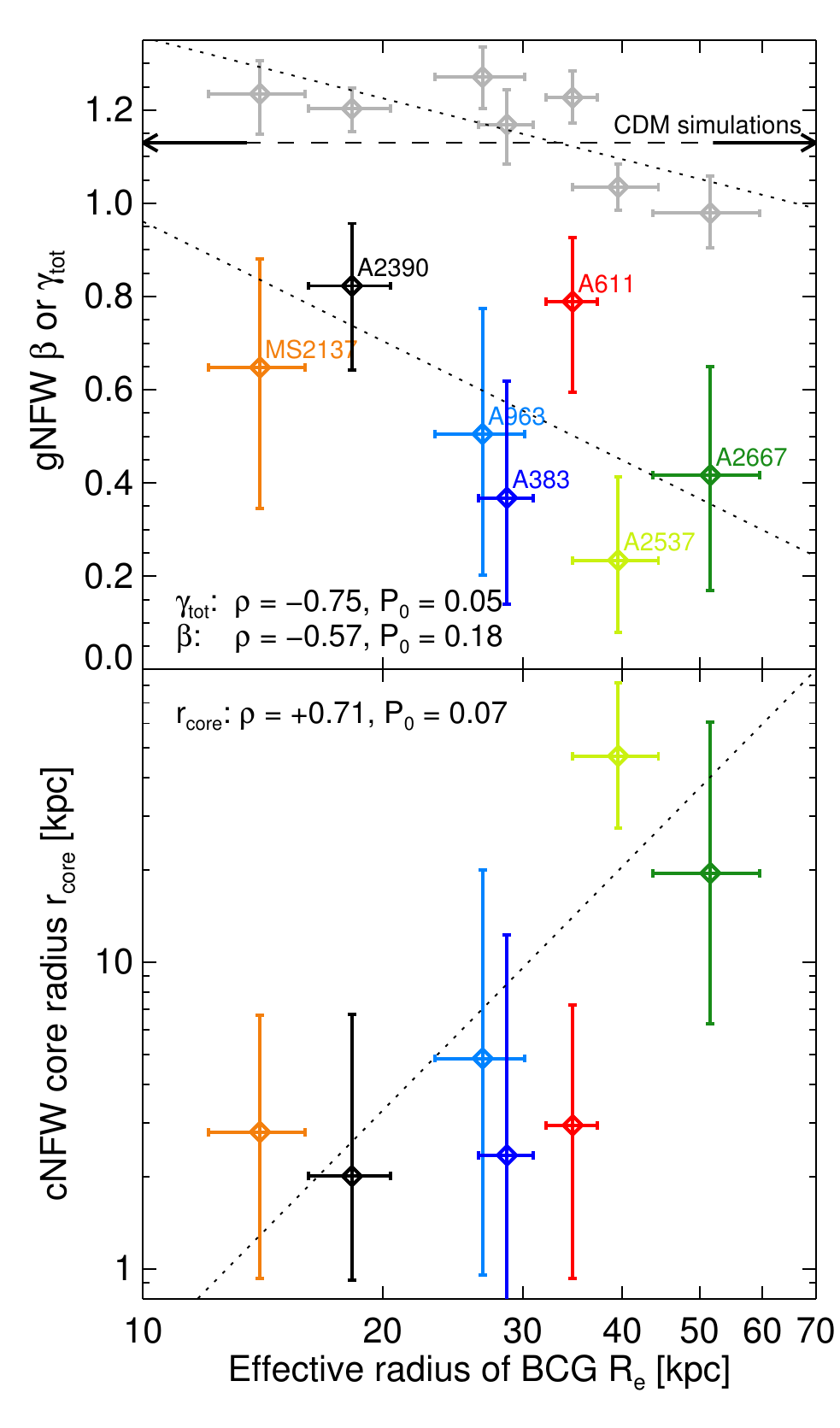}
\caption{Correlation between the size of the BCG and the inner DM profile. \textbf{Top:} Gray points show the total density slope $\gamma_{\textrm{tot}}$ presented in \PaperI; this is measured over $r/r_{200} = 0.003-0.03$ and is not an asymptotic slope. The dashed horizontal line shows the mean slope measured in CDM-only cluster simulations \citep{Gao12} over the same interval. Colored points denote the asymptotic DM density slope $\beta$ measured in the gNFW models. Dotted lines show least-squares linear fits. The Spearman rank correlation coefficient $\rho$ and the corresponding two-sided $P_0$-value are listed. \textbf{Bottom:} The core radii $r_{\textrm{core}}$ of the cNFW models are shown, again indicating a correlation with $R_e$.\label{fig:bcg_halo}}
\end{figure}

We can also ask whether there is evidence for intrinsic variation in the inner DM profiles. This can be quantified by assuming that the parent distributions of $\beta$ and $\log r_{\textrm{core}}$ are Gaussian, and using the method described in Section~\ref{sec:massscale} to infer its dispersion. We find some evidence for intrinsic scatter with $\sigma_{\beta} = 0.22^{+0.15}_{-0.11}$ and $\sigma_{\log r_{\textrm{core}}} = 0.57^{+0.33}_{-0.21}$. Its statistical significance can be assessed with the $\Delta P$ statistic (Equation~\ref{eqn:DeltaP}): we derive $\Delta P = 1.5$ and 2.6 for $\beta$ and $\log r_{\textrm{core}}$, respectively. This indicates a $\simeq 2\sigma$ preference for the presence of intrinsic scatter in the inner DM profile shape. While we have focused on relaxed clusters, we expect this variation would increase if a broader sample of clusters that includes recent mergers were considered.

A possible physical origin of this scatter is illustrated in Figure~\ref{fig:bcg_halo}. Gray points in the top panel show the total density slope $\gamma_{\textrm{tot}}$. As described in \PaperI, these show mild scatter around the mean slope measured in CDM-only simulations \citep[dashed line,][]{Gao12} over the same radial interval ($r/r_{200}=0.003-0.03$). Here we see signs of a correlation with the size of the BCG, with more extended BCGs corresponding to shallower total slopes. The effect on the DM slope (colored points) appears stronger: larger BCGs are hosted by clusters with shallower DM slopes $\beta$, or equivalently larger core radii $r_{\textrm{core}}$ (bottom panel). Such a correlation is necessary for the dark and stellar mass to combine to a similar total density profile. The significance can be assessed using the Spearman rank correlation test. We find a probabilities $P_0 = 0.18$ and 0.07 of obtaining an equally strong correlation between $R_e$ and $\beta$ or $r_{\textrm{core}}$, respectively, in the null hypothesis of uncorrelated data (see caption to Figure~\ref{fig:bcg_halo}).

Figure~\ref{fig:bcg_halo} suggests that the DM profile in the cluster core is connected to the build-up of stars in the BCG. We return to this point in Section~\ref{sec:discussion} and discuss physical scenarios that may explain this. Although the correlations with $R_e$ are most convincing, they are not unique: we find correlations between $\beta$ or $r_{\textrm{core}}$ and the stellar mass or luminosity with nearly equal statistical significance. There is no sign of a correlation with the virial mass $M_{200}$ ($\rho = 0.11$ and 0.04 for the gNFW and cNFW models; see caption to Figure~\ref{fig:bcg_halo}).\footnote{Interestingly, the reverse seems to hold for $\gamma_{\textrm{tot}}$: there is no sign of a correlation with the stellar mass or luminosity, but a possible correlation with $M_{200}$ ($\rho = -0.68$, $P_0 = 0.09$). The latter may simply be because the radial range over which $\gamma_{\textrm{tot}}$ is measured is proportional to $r_{200}$.}

We emphasize that it is preferable to compare directly to the physical density profiles (Figure~\ref{fig:dens}) when possible, rather than only marginalized distributions for $\beta$.  These results do not imply, for example, that a CDM density profile should be modified simply by maintaining the same $r_s$ and changing $\beta = 1$ to $\beta = 0.5$. Rather, $r_s$ also shifts in our fits such that significant changes in $\rho_{\textrm{DM}}$ are kept within $r \lesssim 30$~kpc. This degeneracy is simply a result of the gNFW parameterization. 

\subsection{Systematic uncertainties}
\label{sec:betasys}

A full discussion of the systematic uncertainties affecting our analysis was presented in \PaperI, Section 9.3 (see also \citealt{Sand04}). In the following, we review the most important effects and estimate their impact on $\alpha_{\textrm{SPS}}$ and the inner DM halo parameters $\beta$ and $b$.

One of the main sources of systematic uncertainty is our use of spherical dynamical models based on isotropic velocity dispersion tensors. As discussed in \PaperI~(Section 9.3), this is a good approximation for luminous, non-rotating giant ellipticals in their central regions \citep[e.g.,][]{Gerhard01,Cappellari07}. Nonetheless, individual galaxies can exhibit mild anisotropy with $|\beta_{\textrm{aniso}}| = |1 - \sigma_{\theta}^2 / \sigma_r^2| \approx 0.2$, and the population as a whole also may be slightly radially biased. To estimate the impact this has on our analysis, we repeated the dynamical analysis taking a constant anisotropy parameter $\beta_{\textrm{aniso}} = \pm 0.2$. Arrows in Figure~\ref{fig:alphaIMF} show that individual clusters may shift by $\Delta \log \ML = -0.16$ ($\beta_{\textrm{aniso}} = +0.2$) or $\Delta \log \ML = +0.10$ ($\beta_{\textrm{aniso}} = -0.2$). Since this bias may be correlated among the BCGs, we consider these as systematic uncertainties in the mean: $\langle \log \alpha_{\textrm{SPS}} \rangle = 0.27 \pm 0.05 {}^{+0.10}_{-0.16}$. We note that the effects of anisotropy are larger here than for studies of field elliptical lenses \citep[e.g.,][]{Auger10}, since the latter do not resolve kinematics well within $R_e$ where the impact of anisotropy on the l.o.s. velocity dispersion is largest.

Uncertainties in the orbital distribution have a milder effect on the parameters describing inner DM profile. If we adopt the same prior in $\langle \log \alpha_{\textrm{SPS}} \rangle$, taking $\beta_{\textrm{aniso}} = \pm 0.2$ leads to systematic shifts of $\Delta \langle \beta \rangle = \pm 0.13$ and $\Delta \langle \log r_{\textrm{core}} \rangle \approx -0.18$ (Table~\ref{tab:beta}). If we instead shift the prior on $\langle \log \alpha_{\textrm{SPS}} \rangle$ to match the results obtained with the corresponding $\beta_{\textrm{aniso}}$, we find $\Delta \langle \beta \rangle = +0.11, -0.02$ and $\Delta \langle \log r_{\textrm{core}} \rangle = -0.21, +0.08$. Based on these results, we estimate systematic uncertainties of $\Delta \langle\beta\rangle = \pm 0.13$ and $\Delta \langle \log r_{\textrm{core}} \rangle = -0.2, +0.1$ due to the orbital anisotropy.

We note that the clusters with the lowest inferred $\alpha_{\textrm{SPS}}$ in Figure~\ref{fig:alphaIMF} (MS2137 and A611) are those with the highest halo concentration parameters (\PaperI, Section 10). These clusters have NFW-like \emph{total} density profiles down to unusually small radii, with very weak steeping on small scales. In view of the similarity of $\alpha_{\textrm{SPS}}$ among the other five clusters and the agreement with independent results discussed in Section~\ref{sec:MLcompare}, a likely explanation is that some of the stellar mass is effectively counted in the halo when $\ML$ is allowed to vary freely from cluster to cluster. Nevertheless, omitting MS2137 and A611 would shift $\langle\log \alpha_{\textrm{SPS}}\rangle$ by only $+0.02$.
In this respect our results are encouragingly robust. This highlights the utility of the ensemble of clusters as a robust constraint on $\ML$.


L.o.s. ellipticity in the cluster halo can complicate the coupling of lensing and dynamical mass measurements, since lensing measures the mass contained in cylinders, while dynamical and X-ray measurements nearly measure the spherically averaged mass distribution. The close agreement between lensing- and X-ray-based mass measurements shows that this is not a major effect in our sample; the only exception is A383, in which the l.o.s.~shape is explicitly accounted for in our analysis (\PaperI, Section 8.1, and \citealt{N11}). Specifically, the mean ratio of spherical mass measures $\langle M_{\textrm{X}} / M_{\textrm{lens}}\rangle = 1.1$ at $r \simeq 60$~kpc, the typical Einstein radius in our sample (\PaperI, Section 8). This could be explained by a mean elongation of the cluster halos along the l.o.s.~with ellipticity $\langle q - 1 \rangle \approx 0.1 - 0.2$ (although, as described in \PaperI, $\langle M_{\textrm{X}} / M_{\textrm{lens}}\rangle$ and thus $\langle q \rangle$ are actually consistent with unity within the systematic uncertainties). Based on our study of A383, we estimate that a mean l.o.s.~ellipticity of this magnitude would cause systematic shifts of $\Delta \langle \beta \rangle \approx 0.06$ and $\Delta \langle \log r_{\textrm{core}} \rangle \approx -0.1$.

Combining the effects of l.o.s.~ellipticity and orbital anisotropy in quadrature, we arrive at final measurements $\langle \beta \rangle = 0.50 \pm 0.13 {}^{+0.14}_{-0.13}$ and $\langle \log r_{\textrm{core}} \rangle = 1.14 \pm 0.13 {}^{+0.14}_{-0.22}$ including random and systematic error estimates. Naturally, variations in orbital anisotropy or l.o.s.~ellipticity could cause larger shifts on a cluster-by-cluster basis. Such effects could decrease the intrinsic scatter in $\beta$ and $r_{\textrm{core}}$ that we infer, but they would have to be correlated with the size or mass of the BCG (Figure~\ref{fig:bcg_halo}). While we have argued that our method of deriving a common value of $\alpha_{\textrm{SPS}}$ is superior, we note that marginalizing over $\MLV$ separately in each cluster as in our earlier papers would shift the mean $\langle \beta \rangle$ by $<1\sigma$ (see ``Separate $\alpha_{\textrm{SPS}}$'' Table~\ref{tab:beta}).

In Paper~I we evaluated the effect of varying the positional uncertainty $\sigma_{\textrm{pos}}$ in the strong lensing analysis. In the context of this paper, we find a mean shift of $\Delta \beta = -0.1$ when taking $\sigma_{\textrm{pos}} = 0\farcs3$ rather than our fiducial $\sigma_{\textrm{pos}} = 0\farcs5$, while $\Delta \beta = +0.1$ when $\sigma_{\textrm{pos}} = 1\farcs0$ (although this choice is strongly disfavored by the Bayesian evidence; see Section 7.2 of Paper I). There is no significant dependence of $\log \alpha_{\textrm{SPS}}$ on $\sigma_{\textrm{pos}}$.

Finally, we recall evidence presented in \PaperI~that A2537 is a possible l.o.s.~merger. Such an alignment could produce a spuriously shallow DM profile in a lensing analysis, and A2537 indeed has the shallowest slope in our sample. However, Figure~\ref{fig:bcg_halo} provides another explanation: A2537 has the second-largest BCG in the sample. Thus, it does not appear that our results for A2537 are exceptional. Nevertheless, recognizing its unique nature in our sample, we note that excluding A2537 yields $\langle \beta \rangle = 0.69^{+0.10}_{-0.14}$ and $\log r_{\textrm{core}} = 0.59^{+0.26}_{-0.37}$, which does not change our main conclusions.

\section{Comparison to previous results}
\label{sec:compare}

\subsection{Stellar mass-to-light ratio}
\label{sec:MLcompare}

These results on the inner DM profile are informed by the common stellar mass normalization that we infer, so it is important to compare this result to other measurements to assess its reliability (see also \citealt{Cappellari12c} for a recent review). As shown in Section~\ref{sec:massscale}, we find $\log \alpha_{\textrm{SPS}} = 0.27 \pm 0.05$ for isotropic orbits, with a corresponding $\MLV = 4.1 \pm 0.5$ and $\MLB = 5.3 \pm 0.6$ at the median $\MLVSPS$ and $\MLBSPS$. When comparing mass-to-light ratios at different redshifts, it is essential to account for luminosity evolution. Where necessary, we evolve samples as $d \log \MLV / dz = -0.64$ \citep{Treu01}. We note that the $\simeq0.05$~dex systematic uncertainty in $\MLVSPS$ (\PaperI, Section 5.2) is relevant only for the interpretation of $\MLV$ in terms of stellar populations, but it does \emph{not} affect the stellar mass and so has no effect on the derived mass profiles.

Discussion of $\ML$ is often tied to the IMF. This is because the unknown IMF is the dominant source of uncertainty in the absolute mass scale for SPS models, especially for old galaxies \citep[e.g.,][]{Bell01,Bundy05,Cappellari06,Auger09,Grillo09,Stott10}. If interpreted as a difference in IMF, our measured $\alpha_{\textrm{SPS}}$ indicates a normalization consistent with that of the \citet{Salpeter55} IMF, which has $\log M_{*,\textrm{Salp}}/M_{*,\textrm{Chab}} = 0.25$ when extended over $0.1-100 \msol$.\footnote{While \citet{Salpeter55} did not measure the mass function down to $0.1 \msol$, this is the common meaning of a ``Salpeter'' IMF in extragalactic studies.}

Several other studies have used lensing and stellar dynamics to probe massive field and group ellipticals. \cite{Auger10} study the SLACS samples of early-type lenses using strong and weak lensing and stellar kinematics \citep[see also][]{Gavazzi07,Treu10}. Assuming an NFW halo, they infer $\log \alpha_{\textrm{SPS}} = 0.28 \pm 0.03$ at $M_* = 10^{11} \msol$.\footnote{Their $\alpha_{\textrm{IMF}}$ is defined relative to a Salpeter IMF and so differs from our definition by 0.25~dex.} Assuming an adiabatically-contracted halo lowers this value by $0.11-0.14$, i.e., still heavier than a Chabrier IMF. They infer an intrinsic scatter of $<0.09$~dex in $\log \alpha_{\textrm{SPS}}$ within their sample of $\sigma \gtrsim 200$~km~s${}^{-1}$ lenses \citep{Treu10}. \citet{Lagattuta10} study ellipticals at slightly higher redshift using strong and weak lensing. Evolving their $\ML$ from $\langle z \rangle \approx 0.6$ to our $\langle z\rangle=0.25$ yields $\MLV = 4.7 \pm 0.7$, consistent with our results. Both of these works assume an NFW halo and a mass--concentration relation that follows theoretical expectations (i.e., a one-parameter halo). Our models include much more general halos, and the BCGs are much more DM-dominated. Thus, the uncertainty in $\ML$ on an object-by-object basis is larger; nonetheless, the ensemble averages agree well. \citet{Sonnenfeld12} studied a rare early-type lens that presents two Einstein rings, which allowed them also to relax assumptions on the DM profile. They find $\alpha_{\textrm{SPS}} = 0.30 \pm 0.09$ in our notation (see also \citealt{Spiniello11}). \citet{Zitrin09} took advantage of the unusually flat surface density profile in the lensing cluster MACS~J1149.5+2223 ($z = 0.544$), which offers a clean subtraction of the dark halo to isolate the mass of the BCG. They estimate $\MLB \approx 4.5 \pm 1$ ($\approx 7 \pm 2$ if evolved to our $\langle z \rangle = 0.25$).

Other studies have used integral field spectroscopy to construct detailed dynamical models of local ellipticals. \citet{Cappellari12,Cappellari12c,Cappellari12b} discuss the ATLAS${}^{\textrm{3D}}$ sample of early-type galaxies. At the highest velocity dispersions present, they infer $\log \alpha_{\textrm{SPS}} = 0.25$ (\citealt{Cappellari12c}, Figure 9, converted to our definition of $\alpha_{\textrm{SPS}}$). Interestingly, there appears to be little or no intrinsic scatter in $\alpha_{\textrm{SPS}}$ at $\sigma_e \gtrsim 250$~km~s${}^{-1}$, nearly at the lower limit of our BCGs, although only a handful of such objects are present in their sample. Along with the tightness of the $M/L - \sigma_e$ relation at high $\sigma_e$ \citep{Cappellari12b}, this supports our claim that $\alpha_{\textrm{SPS}}$ should be nearly constant within our sample of BCGs. \citet{McConnell11} studied the BCG of A2162 using long-slit kinematics and integral field spectroscopy with adaptive optics, finding $\MLR = 4.6^{+0.3}_{-0.7}$ in their ``maximum halo'' solution. For comparison, our result evolved to $z = 0$ is $\MLR = 4.1 \pm 0.5$.

Finally, the IMF in early-type galaxies has recently been studied using detailed spectral synthesis models that take advantage of surface gravity-sensitive stellar absorption lines. In very high-quality spectra, these constrain the abundance of low-mass dwarfs that contribute much to the stellar mass but very little to the integrated light. Although the degree of scatter remains unclear, these studies suggest that a Salpeter-like IMF -- or possibly even heavier -- is typical in high-dispersion ellipticals \citep{vanDokkum10,vanDokkum12,Conroy12,Smith12}. 

In summary, our measurements are consistent with a variety of other recent works indicating a heavy (Salpeter-like) $\ML$ in massive early-type galaxies. Encouragingly, studies based on completely independent techniques are beginning to converge on the same results.

\subsection{The total inner density slope}

When comparing results on the inner density profiles of clusters, it is essential to understand the \emph{radial range} that is being fit and whether the \emph{total} density profile or that of the \emph{dark matter} is being considered. This distinction is most important at radii $\lesssim 30$~kpc where the BCG contributes noticeably to the total mass. In \PaperI~we showed that the \emph{total} density profiles in our sample are consistent with CDM-only simulations down to $r \simeq 5-10$~kpc. The mean total density slope $\langle \gamma_{\textrm{tot}} \rangle = 1.16 \pm 0.05 {}^{+0.05}_{-0.07}$ was precisely measured over $r/r_{200} = 0.003-0.03$ and found to be consistent with collisionless CDM-only simulations, which have $\langle \gamma_{\textrm{tot}} \rangle = 1.13$ (\PaperI, Section 9). Note that $\gamma_{\textrm{tot}}$ is measured over a specific radial interval and is distinct from the asymptotic inner slopes of gNFW models, which we denote $\beta_{\textrm{tot}}$ and $\beta_{\textrm{DM}}$ in the following.

Most observational studies have focused on the total density profile. \citet{Umetsu11} stacked density profiles for four clusters with high-quality lensing data and found that $\beta_{\textrm{tot}} = 0.89^{+0.27}_{-0.39}$, with the inner 40~kpc/$h$ excluded from their fit. \citet{Morandi11} measured $\beta_{\textrm{tot}} = 0.90 \pm 0.05$ in A1689, excluding the inner 30 kpc, and \citet{Coe10} also found that the total mass distribution is NFW-like. Using imaging from the CLASH survey \citep{CLASH}, \citet{Umetsu12_J1206} and \citet{Zitrin11} derived $\beta_{\textrm{tot}} = 0.96^{+0.31}_{-0.49}$ (their ``method 7'') and $\beta_{\textrm{tot}} = 1.08 \pm 0.07$ in MACS~J1206.2-0847 and A383, respectively. These lensing results are consistent with our claims that the \emph{total} density profile is NFW-like at $r \gtrsim 5-10$~kpc.

\citet{Morandi12} use lensing and X-ray data to derive a total slope $\beta_{\textrm{tot}} = 1.02 \pm 0.06$ in A383 and contrast this with our earlier finding that $\beta_{\textrm{DM}} = 0.59^{+0.30}_{-0.35}$ in the same cluster \citep{N11}.\footnote{The present measurement of $\beta$ in A383 (Table~\ref{tab:beta}) is slightly shallower, but consistent with, \citet{N11} due to our new joint constraint on $\alpha_{\textrm{SPS}}$.} These results are not inconsistent. Figure~\ref{fig:dens} shows that the DM profile we infer in A383 becomes shallower than an NFW model only at $r \lesssim 30$~kpc. These scales are excluded by Morandi \& Limousin in their fits precisely because of the uncertainty in the BCG stellar mass that we have addressed using stellar kinematics. At $r \gtrsim 30$~kpc the total density profile in our models -- nearly equal to that of the DM -- is NFW-like.

\subsection{The dark matter inner density slope}
\label{sec:dmresults}

Among the main scientific goals of studying the inner regions of clusters are testing predictions of the collisionless CDM paradigm, and understanding the formation of the central galaxy and its impact on the DM halo. Thus, although precise and robust measurements of the total density profile are very valuable, for these goals it is clearly important to understand how much of this mass is DM and how much is baryonic. Over the past decade, we have been developing tools to perform this separation \citep{Sand02,Sand04,Sand08,N09,N11}. The history of this progress was described in Section~\ref{sec:intro}.

\cite{Sand04} measured a mean $\langle \beta_{\textrm{tot}} \rangle = 0.52 \pm 0.05$ in a sample of six clusters. We have improved on this earlier work in many ways: through the use of elliptical lens models, the addition of weak lensing data, the incorporation of multiple strongly lensed sources (usually located at different redshifts), the comparison with X-ray results to quantify l.o.s. effects, the deeper spectroscopic observations of the BCGs that have yielded more precise and radially-extended kinematic profiles, and through joint constraints on the stellar mass scale $\alpha_{\textrm{SPS}}$. This work has essentially confirmed our initial findings, with the present value $\langle \beta_{\textrm{DM}} \rangle = 0.50 \pm 0.10 {}^{+0.14}_{-0.13}$ consistent with \citet{Sand04}. (The smaller error bars quoted in the latter work are due to the more restrictive model assumptions, particularly a fixed scale radius $r_s$.)

Four of the clusters in the present sample have been previously studied in our earlier papers. In general our results for MS2137 and A963 are consistent with \citet{Sand04,Sand08} within their uncertainties, although the present measurements supercede earlier ones due to the improvements described above. Our analysis of A383 is consistent with \citet{N11}. The results presented here for A611, on the other hand, are significantly different from \citet{N09}: we find $\beta = 0.79^{+0.14}_{-0.19}$, rather than $\beta < 0.3$ (68\% confidence). This is attributable to two changes in the data: a revised spectroscopic redshift for a multiply imaged galaxy, and improved stellar kinematic measurements (see Sections~4.4 and 6.4 in \PaperI). 

As we have shown, it is difficult to separate the BCG and DM profiles with lensing alone due to the low density (or lack) of constraints near the center. Only in clusters with exceptional lensing configurations is this feasible. An interesting such case is A1703, which presents an unusual quad image close to the BCG. \citet{Limousin08} and \citet{Richard09} performed a two-component fit -- a gNFW halo and BCG stars following light, as in this work -- and derive $\beta_{\textrm{DM}} = 0.92^{+0.05}_{-0.04}$. (See \citealt{Oguri09} for a consistent result with a much larger error bar.) This may not be inconsistent with our findings, since two clusters in our sample prefer a similar slope (A611 and A2390, see Figure~\ref{fig:DMparams}), and there may be scatter from cluster to cluster.\footnote{\citet{Limousin08} imposed a tight prior on the BCG stellar mass derived from SPS fits, but did not consider uncertainty from the IMF. Their SPS estimates are quite high: $\MLBSPS \approx 11$, whereas we find $\MLBSPS = 3.0$ from fitting the SDSS photometry to this BCG, also using a Chabrier IMF. Adjusting the latter to our preferred $\alpha_{\textrm{SPS}} = 0.27$ yields $\MLB =  5.7$, which agrees with the estimate $\MLB \approx 6$ by \citet{Zitrin10} in this cluster.} \citet{Zitrin10} found that the \emph{total} density profile in A1703 is well-fit by an NFW model.

X-ray studies of two nearby clusters (A2589 and A2029) have also shown that the total density follows an NFW profile down to $\approx 0.002-0.01 r_{\textrm{vir}}$ \citep{Lewis03,Zappacosta06}. The latter authors noted that for any reasonable $\ML$, this implies a shallower DM profile in the central regions where the stellar mass is significant. Their finding agrees well with our work, which has quantified the split between stars and DM. \cite{Schmidt07} studied a large sample of distant X-ray clusters. By assuming a typical BCG stellar mass,  they estimated $\langle\beta_{\textrm{DM}}\rangle = 0.88 \pm 0.29$ (95\% CL). Often the inner $\simeq 40$~kpc must be excluded from their analysis, making a direct comparison difficult.

\begin{figure}
\centering
\includegraphics[width=0.98\linewidth]{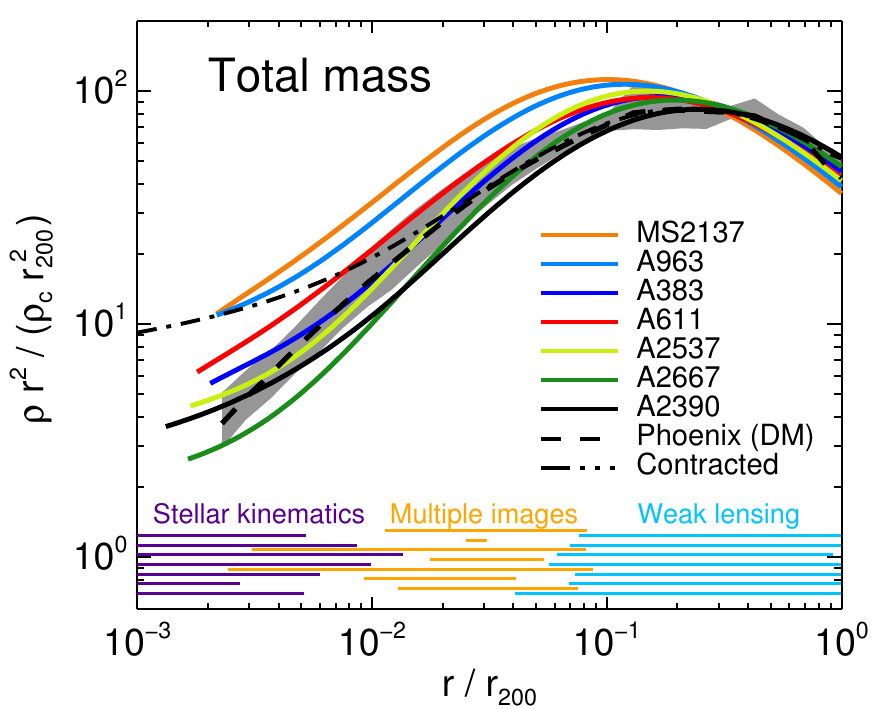} \\
\includegraphics[width=0.98\linewidth]{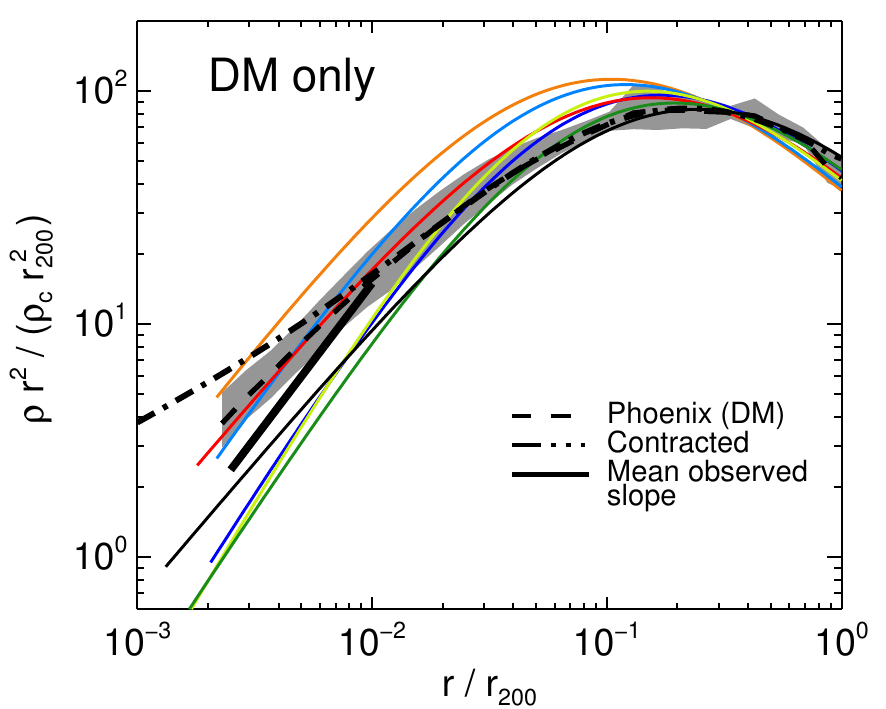}
\caption{\textbf{Top:} \emph{Total} density profiles, including baryons and DM, for our sample are overlaid on CDM-only simulations of massive clusters \citep[][dashed line, with grey band indicating the full range of the simulated clusters; see \PaperI, Section 10]{Gao12}. The dot-dashed line shows a system in which an NFW halo with concentration $c_{200} = 4.5$ is altered using the modified adiabatic contraction model of \citet{Gnedin11}. Parameters of $A_0 = 1.5, w_0 = 0.85$ were used, with the BCG described by a \citet{Jaffe83} profile with scale length $r_J / r_{200} = 0.02$ and mass fraction $M_* / M_{200} = 0.002$, which are representative of our sample. The radial extent of the data is indicated at the bottom of the panel. \textbf{Bottom:} As in the top panel, but showing DM only. (The Phoenix simulations thus do not change.) Note that CDM halos match the observed total density profiles better than those of DM alone. The inclusion of halo contraction (dot-dashed line) only exacerbates the difference with the mean observed DM slope (thick black segment).\label{fig:gaoplot}}
\end{figure}

\section{Discussion and Conclusions}
\label{sec:discussion}

By combining strong lensing, weak lensing, and stellar kinematic observations that extend from $\simeq 3$~kpc to beyond the virial radius, thus spanning the baryon- to DM-dominated regimes, we constrained flexible, physically motivated models of the dark and stellar mass distributions in seven massive, relaxed galaxy clusters. As discussed extensively in \PaperI, the density profiles of stars and DM sum to produce a slope close to \emph{CDM-only} simulations, at least outside the very central $\approx 5-10$~kpc where stars strongly dominate the mass. In this paper we isolated the dark and stellar density profiles to quantify the behavior of the DM on small scales, finding a mean asymptotic inner power law slope of $\langle \beta \rangle = 0.50 \pm 0.13 {}^{+0.14}_{-0.13}$, or equivalently a mean DM core radius $\langle \log r_{\textrm{core}} \rangle = 1.14 \pm 0.13 {}^{+0.14}_{-0.22}$. We also presented evidence for possible variation in the inner DM profile from cluster to cluster, which correlates with the size and mass of the BCG (Figure~\ref{fig:bcg_halo}). This suggests a connection between the DM profile in cluster cores and the assembly of stars in the BCG.

The conclusion that the inner DM profile is shallower than that of pure CDM halos is fully consistent with our previous claims \citep{Sand02,Sand04,Sand08,N09,N11}. We have improved upon these earlier works by collecting improved data for a larger sample of clusters and refining our analysis techniques, as discussed in Section~\ref{sec:dmresults}. A particular advance enabled by this enlarged, improved sample was a joint constraint on the stellar mass-to-light ratio $\ML$ of the BCGs in our sample, which we found to be elevated by $\langle \log \alpha_{\textrm{SPS}} \rangle = 0.27 \pm 0.05 {}^{+0.10}_{-0.16}$~dex relative to fits to SPS models that assume a Chabrier IMF. Our measurements are instead consistent with a Salpeter IMF (or any equivalently ``heavy'' IMF). As reviewed in Section~\ref{sec:MLcompare}, this agrees with recent, independent studies of massive, early-type galaxies based on lensing, dynamics, and detailed spectroscopy \citep{Treu10,Auger10,Cappellari12,Cappellari12c,vanDokkum10,vanDokkum12,Conroy12}. These rapid developments in our understanding of stellar populations promise significant advances in disentangling the distributions of dark and baryonic mass across a range of systems.

Figure~\ref{fig:gaoplot} compares our measurements to high-resolution CDM cluster simulations from the Phoenix project \citep{Gao12}, clearly demonstrating these DM-only simulations are a better match to the total density profile than that of the DM alone. In assessing the role of baryons on their host halos, much of the focus of the theoretical literature has been on the halo contraction \citep[e.g.,][]{Blumenthal86,Gnedin04,Gnedin11} expected to result from a central dissipative build-up of baryons. The dot-dashed lines in Figure~\ref{fig:gaoplot} show the effect of applying the modified adiabatic contraction model of \citet{Gnedin11} to an NFW halo and BCG with parameters typical of our sample and of the Phoenix simulations (see details in caption). As expected, the DM profile steepens (bottom panel), only worsening the disagreement with our observations. It has been argued that this increase in central DM density from adiabatic contraction will boost the gamma-ray flux from DM annihilation in clusters \citep{Ando12}.\footnote{In any case, the highly uncertain contribution from subhalos may dominate this signal \citep[e.g.,][]{Gao12susy}.} However, our results suggest that adiabatic contraction is not the main process that sets the density profile and that the net effect on the halo is actually opposite to predictions. (As discussed in \PaperI, this does not necessarily imply that the same theory cannot make valid predictions at the galaxy scale, where the star formation efficiency and assembly history are very different.)

A possible formation scenario is that early, dissipative star formation in the main BCG progenitor creates a steep total density slope in the inner $\simeq 5-10$~kpc, where stars dominate the mass. This size scale is indeed similar to the observed sizes of very massive galaxies at $z \gtrsim 2.5$ \citep[e.g.,][]{Trujillo06,vanDokkum08,Newman12}. Subsequent assembly of the extended stellar envelope of the BCG -- thought to be dominated by low-mass, dry accretion of satellites \citep[e.g.,][]{Naab09,Laporte12} -- then mostly replaces the DM already in place with satellite material, roughly maintaining the density.

Controlled simulations have indeed shown that dynamical friction between infalling satellites and the DM halo can heat the cusp and reduce the central DM density \citep[e.g.,][]{ElZant01,ElZant04,Nipoti04,Jardel09,Cole11}. This process is dissipationless, since the orbital energy lost by the satellites is transferred to the halo, and thus contrasts with the AC picture, in which the baryons' energy is radiated away \citep{Lackner10}. A connection between the assembled stellar mass and the central DM density is naturally expected. Indeed, \citet{Nipoti04} find an anti-correlation between the amount of stellar mass assembled in the BCG and the inner DM density slope $\beta$, similar to our observations (Figure~\ref{fig:bcg_halo}; we note the satellites in their simulations included no DM). \citet{DelPopolo12} discusses a similar anti-correlation arising in their analytic models for the same physical reason, with higher central baryon fractions corresponding to shallower DM density cusps. The strength of the dynamical friction effect depends on the density of the satellites and their resistance to stripping. \citet{Laporte12} showed that when a stellar mass--size relation in line with $z \gtrsim 2$ observations is imposed in their simulations (offset by $3-5\times$ in size from the local relation), the central DM cusp is flattened to $\beta \simeq 0.3 - 0.7$, comparable to our observations. It is important to realize that numerical experiments investigating this dynamical effect have generally lacked a fully realistic and consistent treatment of the satellites, so improved simulations are needed. Nonetheless, the current results are promising.

Until the last few years, full hydrodynamical, cosmological cluster simulations that include cooling, star formation, and feedback did not produce shallow DM cusps or cores, which probably reflected overcooling effects. \cite{Mead10} and \citet{Martizzi12} showed that the inclusion of active galactic nucleus (AGN) feedback greatly improves this situation (see discussion and references in \PaperI, Section 10) and may also play a key role in lowering the central DM density. Understanding the  impact that gas cooling, dynamical friction from stellar "clumps," and AGN feedback have on the small-scale DM distribution is an important avenue for future simulations, and the data we have presented provide strong constraints.

In addition to the effect of baryons on the halo, various DM particle scenarios have also been proposed to reduce tension between CDM and observations on small scales, including the ``missing satellites'' problem and evidence for central DM cores or shallow cusps (for a recent review, see \citealt{Primack09}). These include warm sterile neutrinos at the $\sim\textrm{keV}$ scale \citep[e.g.,][]{Abazajian01,Boyarsky09,Maccio12,Menci12}, ``fuzzy'' CDM composed from an ultralight scalar particle \citep{Hu00,Woo09}, DM produced from early decays \citep{Kaplinghat05}, and DM that itself decays with a long timescale \citep{Peter10}, among many other possibilities. The goal is to preserve the large-scale successes of CDM, while allowing for modifications at higher densities where the detailed properties of the DM particle might manifest.  A scenario for which halo density profiles has been worked out in detail is a self-interacting DM particle \citep{Spergel00,Yoshida00,Dave01}. \citet{Rocha12} and \citet{Peter12} showed that a cross-section $\sigma \sim 0.1$~cm${}^2$~g${}^{-1}$ can produce $\approx 20$~kpc cores in clusters without violating any current constraints, e.g., from the asphericity of cluster cores or the Bullet Cluster \citep{Randall08}. Only the dense central regions of the halo are affected, where scattering can occur within a Hubble time.

These $\approx 20$~kpc core sizes are intriguingly similar to our observations. On the other hand, they are also very similar to the scale of the baryons, i.e., the size of the BCG. It is unclear why the total density profile should then match the shape expected of collisionless CDM. In these scenarios, the core size arises from the microphysics of the DM particle and presumably should not ``know'' about the size of the central galaxy (Figure~\ref{fig:bcg_halo}), for example. Thus, observations of clusters alone cannot provide unambiguous support for alternative DM theories. Global comparisons across a wide range of mass scales (for instance, a cross-section that also produces correct core sizes and densities in dwarf galaxies) remain an essential test for attempts to explain low central halo densities in terms of the DM particle.

\acknowledgments
It is a pleasure to acknowledge helpful conversations with Annika Peter. We thank Liang Gao for providing the Phoenix simulation results. The anonymous referee is thanked for a helpful report. R.S.E.~acknowledges financial support from DOE grant DE-SC0001101. Research support by the Packard Foundation is gratefully acknowledged by T.T. The authors recognize and acknowledge the cultural role and reverence that the summit of Mauna Kea has always had within the indigenous Hawaiian community. We are most fortunate to have the opportunity to conduct observations from this mountain.

\bibliographystyle{apj}
\bibliography{paper2}

\begin{appendix}
In Section~\ref{sec:innerslopes}, we described how posterior probability distributions $P(\beta)$ and $P(\log r_{\textrm{core}})$ are derived for each cluster by weighting the samples in the Markov chains derived in \PaperI. The weights
\begin{equation}
w = \frac{1}{\sqrt{2\pi} \sigma} \exp\left[-\frac{1}{2} \left(\frac{\log \alpha_{\textrm{SPS}} - \langle \log \alpha_{\textrm{SPS}} \rangle}{\sigma}\right)^2\right]
\label{eqn:weights}
\end{equation}
effectively convert a flat prior on $\log \alpha_{\textrm{SPS}}$ (\PaperI, Section 7) to a Gaussian with mean $\langle \log \alpha_{\textrm{SPS}} \rangle = 0.27$ and a dispersion $\sigma = (\sigma_{\alpha}^2 + \sigma_{\textrm{SPS}}^2)^{1/2}$. This dispersion accounts for two sources of error: the uncertainty $\sigma_{\alpha}=0.05$~dex in the global systematic offset $\langle \log \alpha_{\textrm{SPS}} \rangle$ from SPS estimates $\MLVSPS$, and the random photometric uncertainty $\sigma_{\textrm{SPS}}=0.07$~dex in $\MLVSPS$ for each cluster. The first uncertainty is correlated across the entire sample, while the second is not.

Therefore, to obtain constrains on the mean $\langle \beta \rangle$ and $\langle \log r_{\textrm{core}} \rangle$, the probability distributions derived for each cluster in this manner cannot simply be multiplied, since they are not independent. Instead, we calculate the posterior probability of $\langle \beta \rangle$ as
\begin{equation}
P(\langle \beta \rangle) \propto \int P(\langle \beta \rangle | \log \alpha_{\textrm{SPS}}) P(\log \alpha_{\textrm{SPS}})\,d\alpha_{\textrm{SPS}}.
\end{equation}
Here $P(\langle \beta \rangle | \log \alpha_{\textrm{SPS}})$ is the posterior distribution of $\langle \beta \rangle$ at a fixed value of $\log \alpha_{\textrm{SPS}}$. It is obtained by multiplying the probability densities $P(\beta | \log \alpha_{\textrm{SPS}})$ for the seven clusters in our sample, which are each computed with Gaussian weights centered at the fixed value of $\log \alpha_{\textrm{SPS}}$ and a dispersion $\sigma_{\textrm{SPS}}$ (i.e., $\sigma = \sigma_{\textrm{SPS}}$ in Equation~\ref{eqn:weights}; we now account for only the random photometric errors in $\MLVSPS$ since $\log \alpha_{\textrm{SPS}}$ is fixed). $P(\log \alpha_{\textrm{SPS}})$, which represents our constraint on the common stellar mass scale, is simply a Gaussian with mean $\langle \log \alpha_{\textrm{SPS}} \rangle = 0.27$ and dispersion $\sigma_{\alpha} = 0.05$~dex, as derived in Section~\ref{sec:massscale} for isotropic orbits.

We estimate the intrinsic scatter in $\beta$ (Section~\ref{sec:innerslopes}) using the posterior probability densities $P(\beta | \log \alpha_{\textrm{SPS}}=0.27)$ for each cluster. That is, we evaluate the cluster-to-cluster scatter in $\beta$ at a fixed value of $\log \alpha_{\textrm{SPS}}$. All of the above comments apply equally to our study of the cNFW models, simply replacing $\beta$ by $\log r_{\textrm{core}}$.

\end{appendix}
\end{document}